\documentclass[copyright,creativecommons]{eptcs}
 \providecommand{\event}{Gabriel Ciobanu (Ed.): Membrane Computing and \\
Biologically Inspired Process Calculi 2009} 
	\providecommand{\volume}{15}   
	\providecommand{\anno}{2009}   
	\providecommand{\firstpage}{1} 
	\providecommand{\eid}{1}       

\usepackage{latexsym, amssymb}
\usepackage[all,dvips]{xy}
\usepackage{breakurl}

\newcommand{\flt}[1]{\ensuremath{[\![{#1}]\!]}}

\newcommand{\Pow}[1]{\ensuremath{\mathbf{2}^{\!~#1}}}

\newcommand{\Powfin}[1]{\ensuremath{\mathbf{2}^{\!~#1}_{\!~\mathit{fin}}}}

\newcommand{\evrule}[4]{#1 \rightarrow^{#3}_{#4} #2} 

\newcommand{\zero}{{\mathbf 0}}

\newcommand{\trans}[1]{\ensuremath{\,[\/{#1}\/\rangle}\,}

\newcommand{\membtonet}{\mathcal{F}}

\newcommand{\Nat}{{\mathbb N}}

\newcommand{\nplace}[4]{*=(4,4)[o][F-] {#2} 
  \save []+<#3,#4>*\txt{\ensuremath{#1}} \restore}
\newcommand{\zplace}[4]{*=(2,2)[o][F-] {#2} 
  \save []+<#3,#4>*\txt{\ensuremath{#1}} \restore}

\newcommand{\trs}[1]{*=(8,7)[][F-]{#1}}

\newcommand{\iar}{\ar@{*\cir<2pt>{}{.}{}}}

\newcommand{\DE}{\!\!\xymatrix@C=2mm{{}  & \ar@{*\cir<1pt>{}{-}{|}}[l] {}}\!\!}

\newcommand{\pre}[1]{\ensuremath{\!~^\bullet{#1}}}

\newcommand{\post}[1]{\ensuremath{{#1} {^\bullet}}}

\newtheorem{definition}{Definition}
\newtheorem{theorem}{Theorem}
\newtheorem{proposition}{Proposition}

\title{Dependencies and Simultaneity in Membrane Systems}
\author{G. Michele Pinna
\institute{Dipartimento di Matematica e Informatica\\ 
Universit\`a degli Studi di Cagliari\\ 
Cagliari, Italy}
\email{gmpinna@unica.it}
\and
Andrea Saba
\institute{Dipartimento di Matematica e Informatica\\
Universit\`a degli Studi di Cagliari\\ 
Cagliari, Italy}
\email{andrea@sc.unica.it}
}

\def\titlerunning{Dependencies and Simultaneity in Membrane Systems}
\def\authorrunning{G.~M. Pinna \& A. Saba}

\begin{document}

\maketitle

\begin{abstract} 
Membrane system computations proceed in a synchronous fashion: at each step all the 
applicable rules are actually applied. Hence each step \emph{depends} on the previous one.
This coarse view can be refined by looking at the dependencies among rule occurrences, by recording, 
for an object, which was the a rule that \emph{produced} it and subsequently (in a later step),
which was the a rule that \emph{consumed} it. 
In this paper we propose a way to look also at the other main ingredient in membrane system 
computations, namely the simultaneity in the rule applications. 
This is achieved using \emph{zero-safe nets} that allows to \emph{synchronize} transitions, 
i.e., rule occurrences.
Zero-safe nets can be unfolded into occurrence nets in a classical way, and to this unfolding an event structure can be associated. The capability of capturing simultaneity of zero-safe nets is transferred
on the level of event structure by adding a way to express which events occur simultaneously.
\end{abstract}

\section{Introduction}\label{se:introduction}
The study of the dependencies among rule application in membrane systems, introduced in \cite{Paun:EATCS99,Paun:CompwithMemb2000}, have been recently a subject of investigation.
The interest in capturing notions like {\em concurrency} and {\em causality} in membrane systems has arisen (e.g., \cite{Busi:causality,Ciobanu:events}). In fact, as in membrane systems parallelism and concurrency are present, it is worthwhile to understand their relationships with more classical model where these notions are represented. 

Membrane systems are based upon the notion of membrane structure, which
is a structure composed by several membranes, hierarchically embedded in a
main one, called the skin membrane\footnote{A plane representation of a membrane
structure can be given by means of a Venn diagram, without intersected sets and
with a unique superset.}. The membranes delimit regions (compartments) and to each region we associate a (multi)set of objects, described by some symbols over an alphabet, and a
set of evolution rules, which can modify the objects to obtain new objects and possibly send them outside the membrane or to an inner membrane. The various compartments have different tasks, and all together they contribute to accomplishing a more complex one.
The evolution rules are usually applied in a maximally
parallel manner: at each step, all the objects which can evolve should evolve.
If we start from an initial configuration, with
a certain number of objects in certain membranes, and we let the system evolve, we obtain a computational device.
If a computation halts, that is no further evolution rule can be applied, the
result of the computation is defined to be the number of objects in a specified
membrane (or expelled through the skin membrane). If a computation never
halts (i.e., one or more objects can be rewritten forever), then it provides no
output.

In this paper we continue an investigation started in \cite{PiSa:ASC}, where we sudied causality and concurrency in membrane systems with promoters and inhibitors with the aid of zero safe nets and event automata.

The idea of relating membrane systems and nets is not new. 
In \cite{KKR:PetriNetsandMembrane2005,KKR:ProcessforMembrane2006} a comparison with a suitable class of nets, Place/Transition nets with explicit localities, is used to capture the compartment structure of membrane systems. Each locality identifies a distinct set of transitions which may only be executed synchronously, i.e., in a locally maximal concurrent manner; and a notion of process for these nets is developed, with the associated notion of dependency. This is further studied and cast in a more general framework in \cite{KK:TCS2008}. 
We develop a Petri net view of membrane systems which is, differently from other approaches, based on zero safe nets \cite{BM:ZSInfoandCo2000}. 
Zero safeness takes into account the compound nature of the evolution step, based on the application of several rules. As zero safeness allows to synchronize transitions in nets, it seems to be, in our opinion, the correct notion to be used in this setting.
In particular, zero safe nets allow to define in a more clean way what the application of a set of rules in a membrane system is, using zero safe places to represent partial configurations, introduced Busi in \cite{Busi:TCS2007}, which are useful in capturing dependencies among rules applications. 

Concerning causality, 
Busi in \cite{Busi:causality} studied the causal dependencies by introducing a notion of reaction which can be decorated with names, and in \cite{Ciobanu:events} Ciobanu and Lucanu studied which kind of event structure arises from membrane systems. Here the interesting feature is the attempt to identify differently what an \emph{event} is in membrane computing, not as single occurrence of rules but as the computational entity changing the state. 

Here we add to event structure a notion of \emph{simultaneity} to capture, among the concurrent events, which are those that are not only independent but have to happen simultaneously.
Indeed the computation in a membrane system proceeds as a sequence of steps where each step depends totally on the previous ones and within each step all the rules are applied at the same time. 
The Petri net view of a membrane system allow to have a finer description of the causality, identifying which rule actually produces an object used by another rule (in a subsequent step), but also, using the notion of transaction in zero safe nets \cite{BM:TransZSnet}, it possible to characterize the simultaneous occurrence of several rules. Thus to the unfolding of zero safe nets an event structure with simultaneity can be easily associated. 
 
Let us illustrate this idea with a simple example.
Consider the P system
$$(\{a,b,c\}, [_1\ [_2\ ]_2\ ]_1, c, aab, \{r_1 = bc \rightarrow (a,\mathit{here})\}, \{r_2 = a \rightarrow (b,\mathit{out}), r_3 = b \rightarrow (c,\mathit{here})\})$$
with three objects ($a,b$ and $c$), two membranes, two sets of rules, the one associated to the external membrane ($r_1$) transforming the multiset with one occurrence of $b$ and one of $c$ in the multiset with just one $a$, and the set of rules associated to the internal membrane contains two rules: $r_2$ transforming the multiset with one occurrence of $a$ in the multiset with one occurrence of $b$ which is sent to the outer membrane and $r_3$ which transforms $b$ in $c$. The initial state is the multiset with just one occurrence of $c$ in the external membrane and the multiset with two occurrences of $a$ and one of $b$ in the internal one. In the first step two instances of the rule $r_2$ are used together one instance of the rule $r_3$, and the new state is the multiset with two occurrences of $b$ and one of $c$ in the external membrane and the multiset with just one $c$ in the internal one. Now just one instance of the rule $r_1$ can be applied, and the unfolding of the net associated to the  allows to say that $r_1$ depends on one of the instances of $r_2$. Furthermore we are able to characterize the set of events containing the two occurrences of $r_2$ and one of $r_3$ as a set of simultaneous events. 

Simultaneity and concurrency do not coincide in general: simultaneous events are clearly concurrent, but the vice versa does not hold in general. 
Consider the P system, which is similar to the one seen before:
$$(\{a,b,c\}, [_1\ [_2\ ]_2\ ]_1, cc, aa, \{r_1 = bc \rightarrow (a,\mathit{here})\}, \{r_2 = a \rightarrow (b,\mathit{out})\})$$
The two instances of the rule $r_1$ and the two instances of the rule $r_2$ are clearly independent. Each application of the rule $r_2$ create an object $b$ used by the rule $r_1$.
Let us call the instance of these rules as $r_1, \hat{r}_1, r_2$ and $\hat{r}_2$ and assume the the $b$ produced by $\hat{r}_2$ is used by $\hat{r}_1$. Clearly $r_1$ and $\hat{r}_2$ are concurrent (i.e., they are neither causally related nor in conflict) but they are not simultaneous (they occur in different steps).
This motivates the need of a notion of simultaneity in the event structure associated to membrane systems. 
Here we do not put forward any mean to \emph{deduce} simultaneity from other relations, we just observe that the kind of event structures studied in \cite{PP:NEPC, BBCP:rivista} and \cite{Pinna06:FI} can be used to characterize this notion.
In the paper (\cite{PiSa:ASC}) we adopted the collective token semantics approach, arguing that the individual token philosophy is too fine grained for membrane computing, as distinguishing among two occurrence of the same rule to determine causality is somehow deceptive. 
Here we pursue the individual token philosophy as our main contribution is the idea of focussing not only on the causal dependencies but also on the notion of simultaneity among rule occurrences. 
It is, at this stage of our findings, unclear how simultaneity and the collective token philosophy interact, but this will be subject to further investigations.

The paper is organized as follows: in the next section we will fix some notation to be used in the paper and then, in section~\ref{psyst} we review the notion of P system. In section~\ref{sec:net} we present the relevant notions about zero safe nets, occurrence nets and unfolding, and then, in section \ref{sec:ees} we introduce the notion of event structure with simultaneity and relate it to 
prime event structure.
In \ref{sec:psystandnet} we relate membrane systems and nets and in section~\ref{sec:evstruandmemb} we show the event structure semantics for membrane system capturing both dependencies and simultaneity.

\section{Background}\label{background}
With $\Nat$ we denote the set of natural numbers and $\Nat^{+} =
\Nat\setminus\{0\}$, furthermore with $\mathbb{Z}$ we denote the set of positive and negative numbers.
Given a set $S$, a {\em multiset} over $S$ is a function 
$m: S \rightarrow\Nat$. A multiset $m$ over $S$ is \emph{finite} 
iff the set $dom(m) = \{ s \in S \, | \,
m(s) \neq 0\}$ is finite. 
The {\em multiplicity} of an element $s$
in $m$ is given by 
$m(s)$.
The set of multisets of $S$ is denoted by $\mu S$.
A multiset $m$ such that $dom(m) = 
\emptyset$ is called {\em empty} and it is denoted by $\mathbf 0$. 
With $\Pow{S}$ we denote the set of the subsets of $S$. The set of all finite
sets over $S$ is denoted by $\Powfin{S}$.
The cardinality of a multiset is defined as $|m| = \sum_{s\in S}m(s)$.
We write $m \subseteq m'$ if $m(s) \leq m'(s)$ for all $s \in S$, and  $m \subset m'$ if
$m \subseteq m'$ and $m \neq m'$.
The operator $\oplus$ denotes {\em multiset union}: $m \oplus m'(s) =
m(s) + m'(s)$. The operator $\setminus$ denotes {\em multiset difference}:
$m \setminus m'(s) =$ if $m(s)>m'(s)$ then  $m(s)-m'(s)$ else $0$.
The {\em scalar product} of a number 
$j$ with a multiset $m$ is $(j \cdot m)(s) = j \cdot (m(s))$.
If $m \in \mu S$, we denote by $\flt{m}$ the multiset defined as $\flt{m}(a)
 = 1$ if $m(a) > 0$ and $\flt{m}(a) = 0$ otherwise; sometimes $\flt{m}$
 will be confused with the corresponding subset $\{ a \in A \mid
 \flt{m}(a) = 1 \}$ of $A$.
 A {\em multirelation} $f$ from $S$ to $S'$ (often indicated as $f : S \to S'$) 
 is a multiset of $S \times S'$. We
 will limit our attention to finitary multirelations, namely
 multirelations $f$ such that the set $\{ s' \in S' \mid f(s,s') > 0 \}$
 is finite. A multirelation $f$ induces a function
 $\mu f$ from $\mu S$ to $\mu S'$, defined as 
 $\mu f(\sum_{s \in S }n_s
 \cdot s) = \sum_{s' \in S'} \sum_{s \in S} (n_s \cdot f(s,s')) \cdot s'$
 (possibly partial, since infinite coefficients are disallowed). Whenever $f$
 satisfies $f(s,s') \leq 1$ for all $s \in S$ and $s' \in S'$, i.e.,  $f =
 \flt{f}$, we sometimes confuse it with the corresponding
 set-relation and write $f(s,s')$ for $f(s,s') =1$.

The language of \emph{membrane structure}, denoted with $MS$, is a language over $\{[,]\}$ whose strings are defined as follows: (i) $[\ ]\in MS$,
(ii) if $\mu_{1},\dots,\mu_{n} \in MS$, with $n\geq 1$, then $[\mu_{1}\dots\mu_{n}] \in MS$, and
 nothing else is in $MS$.
The same membrane structure can be represented by several equivalent strings (the equivalence being that $\mu_1\mu_2\mu_3\mu_4 \equiv \mu_1\mu_3\mu_2\mu_4$, for $\mu_1\mu_4\in MS$ and $\mu_2, \mu_3\in MS$) hence we assume that the canonical representation of it is given by a tree like structure, i.e., a membrane structure is a rooted tree. 
We call \emph{membrane} each matching pair of parentheses appearing in the membrane structure.
 
Given a membrane structure $\mu$, the number of (nested) membranes is defined as follows: $mem([\ ]) = 1$ and $mem([\mu_1\dots\mu_n]) = \sum_{i=1}^{n} mem(\mu_i) + 1$. The depth of a membrane structure $\mu\neq [\ ]$, i.e., the maximal number of nested membranes, is easily defined as $\mathit{depth}([\ ]) = 1$ and
$\mathit{depth}(\mu) = max\{\mathit{depth}(\mu_i)\ |\ \mu = [\mu_1\dots\mu_n],\ 1\leq i \leq n\}+1$. The depth of the membrane $[\ ]$ is equal to $0$.
Given a membrane $\mu$, to each nested membrane it is possible to associate a unique index (from $1$ to $mem(\mu)$), hence we freely identify a membrane with an index (the convention being that $\mu$ has the number 1, and if $i$ is the index of a membrane $\mu_i$ and $j$ is the index of a nested membrane of $\mu_i$, then $j>i$). Given a membrane $i$, $i\neq 1$, it has a father, which is the membrane $j$, $j\leq i$, such that $\mu_j = [\dots\mu_i\dots]$. The function $\mathit{father}(i)$ returns the index of the father membrane if $i>1$ and it is undefined otherwise (thus the father of the outer membrane does not exist). A membrane $j$ can have children, i.e., membranes $j_1,\dots,j_k$ such that $\mathit{father}(j_1) = \dots = \mathit{father}(j_k) = j$. Hence the function $\mathit{children}$ returns a set of indexes $\mathit{children}(i) = \{j \ | \ \mathit{father}(j) = i\}$.
A membrane structure can be seen not only as a rooted tree, but is often represented as Venn diagram in which any closed space (delimited by a membrane and by the membranes immediately inside) is called a \emph{region} (or compartment).

\section{P systems}\label{psyst} 
In this section we recall the definition of \emph{membrane systems}, also called P systems. 

\begin{definition}
 A \emph{membrane system} over $V$, a finite alphabet of (names of) objects or molecules, is a  
 construct $\Pi = (V,\mu, w_{1}^{0},\dots,w_{n}^{0},R_{1},\dots,R_{n})$ where:
 \begin{itemize}
  \item      $\mu$ is a \emph{membrane structure} with $n$ membranes indexed $1, \dots, n$,
  \item      each $w_i^{0}$ is a multiset over $V$ associated with membrane $i$, and 
  \item      each $R_{i}$ is a finite set of \emph{reaction (or evolution) rules} $r$ associated
      with the membrane $i$, of the form $\evrule{u}{v}{}{}$, where $u$ is a finite
      multisets over $V$, and $v$ is a finite multiset over 
      $V \times (\{here,out\}\cup\{in_j\ |\ \mathit{father}(j)=i\})$, and each rule is such that
      $u\neq \zero$. 
 \end{itemize}
\end{definition}
Given a rule $r$, $u$ is the left hand side of $r$ and
$v$ is the right hand side of $r$.
To ease the notation, given a rule $r = \evrule{u}{v}{}{}$,
with $\pi(v)|_{\alpha}$ we denote the multiset on $V$ obtained from $v$ by considering all the elements with the second component equal to $\alpha$.
In the following we will often omit $(V,\mu, w_{1}^{0},\dots,w_{n}^{0},R_{1},\dots,R_{n})$
when it is possible and no confusion arises, and indicate a membrane system simply with $\Pi$.

A membrane system $\Pi$ evolves from configuration to configuration as a consequence of the application of (multisets of) evolution rules in each region. We start formalizing the notion of configuration of a membrane system.
Following Busi \cite{Busi:causality} and \cite{Busi:TCS2007}, we introduce the notion of partial configuration, which captures the following idea: the state of each membrane is divided in two parts, what can be used (consumed), and what is produced \emph{during} the evolution, pointing out in a clearer way the effects of each rule application. 
With the aid of this notion the evolution of a membrane system is described by two relations: one among partial configurations (micro steps) and another among configurations (macro steps). Clearly the two notions are tightly related, as we will see later.
\begin{definition}
Let $\Pi$ be a  membrane system, then a \emph{configuration} is a tuple $C = (w_1,\dots,w_n)$ where each $w_i$ is a multiset over $V$. $C_0 = (w_{1}^{0},\dots,w_{n}^{0})$ is the \emph{initial} configuration of $\Pi$. The set of configurations of a membrane system is denoted with $\mathit{Conf}_\Pi$.

A \emph{partial configuration} is a tuple $C = ((w_1,\overline{w}_1),\dots,(w_n,\overline{w}_n))$ where each $w_i,\overline{w}_i$ is a multiset over $V$. $C_0 = ((w_{1}^{0},\zero),\dots,(w_{n}^{0},\zero))$ is the \emph{initial} partial configuration of $\Pi$. The set of partial configurations of a membrane system is denoted with $\mathit{PConf}_\Pi$.
\end{definition}
To each configuration $(w_1,\dots,w_n)$ a partial configuration corresponds, namely 
$((w_1,\zero)\dots,(w_n,\zero))$.
A configuration is obtained from a partial one by simply adding, for each pair $(w_i,\overline{w}_i)$, to the left hand side $w_i$, the right hand side $\overline{w}_i$, and then by setting each right hand side to $\zero$) and then forgetting the right hand sides\footnote{This operation is called $\mathit{heated}(\cdot)$ in \cite{Busi:causality}.}. 
Let $\Pi$ be a  membrane system and let 
$((w_1,\overline{w}_1),\dots,(w_n,\overline{w}_n))$ be a partial configuration. Then $\mathit{heated}(((w_1,\overline{w}_1),\dots,(w_n,\overline{w}_n))) = (w_1\oplus\overline{w}_1,\dots,w_n\oplus\overline{w}_n)$. 

We formalize the evolution of a membrane system in a slightly different way with respect to the classical approach. We proceed in two stages: first we define a partial reaction relation (which is basically the microstep relation) describing the effects of the actual application of a rule and then, using this one, we define the evolution of the whole system. 
Let $\Pi = (V,\mu, w_{1}^{0},\dots,w_{n}^{0},R_{1},\dots,R_{n})$ be a  membrane system, then a \emph{vector multi-rule} $\vec{R}$ is the $n$-uple $(\widehat{R}_1,\dots,\widehat{R}_n)$ where $\widehat{R}_i$ is a multiset over $R_i$. The set of vectors multi-rule is denoted by $\mathcal{R}$.

\begin{definition}\label{de:reacpart}
Let $\Pi$ be a  membrane system, and let $\gamma = ((w_1,\overline{w}_1),\dots,(w_n,\overline{w}_n))$ be a partial configuration. Assume there exists an $i$ with $1\leq i\leq n$, and a rule $r = \evrule{u}{v}{}{}\in R_{i}$ such that:
\begin{itemize}
\item $u\subseteq w_i$, 
\item $w'_i = w_i \setminus u$ and 
      $\overline{w}'_i = \overline{w}'_i \oplus \pi(v)|_{\mathit{here}}$,
\item 
      $\forall j\neq i$ $w'_j = w_j$, 
\item for $i\neq 1$, $\overline{w}'_{\mathit{father}(i)} = \overline{w}_{\mathit{father}(i)} \oplus  
      \pi(v)|_\mathit{out}$,
      $\forall j \in \mathit{children}(i)$, 
      $\overline{w}'_j = \overline{w}'_j \oplus \pi(v)|_\mathit{in_j}$, and
\item $\forall j\ j\neq i$ and $j\not\in\mathit{children}(i)$ and $j\neq\mathit{father}(i)$ 
      it holds that $\overline{w}'_j = \overline{w}_j$.
\end{itemize}
Then $\gamma = ((w_1,\overline{w}_1),\dots,(w_n,\overline{w}_n)) \stackrel{\{(r,i)\}}{\longmapsto} ((w'_1,\overline{w}'_1),\dots,(w'_n,\overline{w}'_n)) = \gamma'$. 

The \emph{partial reaction relation}, denoted with $\stackrel{\{(r,i)\}}{\longmapsto}$, is a subset of  $\mathit{PConf}_\Pi \times \mathcal{R} \times \mathit{PConf}_\Pi$.
\end{definition}

With $\gamma\not\longmapsto$ we denote the fact that, for all possible indexes $1\leq i\leq n$, there is no rule that it is applicable at the given partial configuration. 
The transitive closure of this relation, denoted with $\ \stackrel{A}{\longmapsto}{}\!\!^{+}$, is defined as follows: if
$\gamma \stackrel{\vec{R}}{\longmapsto} \gamma'$ and $\gamma' \stackrel{\vec{R}'}{\longmapsto} \gamma''$ then $\gamma \stackrel{A}{\longmapsto} \gamma''$, where $A = \vec{R}\oplus\vec{R}'$ (we use the same symbol to denote the vector sum and the multiset sum).

We formalize the more classical notions of evolution of a membrane system. The maximality is taken into account by the micro step relation. 
We  stress that the order of evolution rules application in the micro step is inessential.

\begin{definition}\label{de:reacrel-ms}
Let $\Pi$ be a  membrane system. The \emph{reaction relation} $\Longrightarrow \subseteq \mathit{Conf}_\Pi \times  \mathcal{R}\times \mathit{Conf}_\Pi$ is defined as follows:
$(w_1,\dots,w_n) \stackrel{\vec{R}}{\Longrightarrow} (w'_1,\dots,w'_n)$ iff there exists a partial configuration $\gamma'$ such that $\gamma' \not\longmapsto$, $((w_1,\zero),\dots,(w_n,\zero))\stackrel{\vec{R}}{\longmapsto}{}\!\!^{+}\gamma'$ and $\mathit{heated}(\gamma')) = (w'_1,\dots,w'_n)$. 
\end{definition}
The transitive and reflexive closure of $\Longrightarrow$ is defined as in the case of $\longmapsto$:
$C \stackrel{\vec{R}}{\Longrightarrow} C'$ and $C' \stackrel{\vec{R}'}{\Longrightarrow} C''$ then $C \stackrel{A}{\Longrightarrow} C''$, where $A = \vec{R}\oplus\vec{R}'$.
We can now formalize the notion of \emph{reachable} configuration. 
\begin{definition}
Let $\Pi$ be a  membrane system, and $C$ be a configuration. Then $C$ is \emph{reachable} iff $C_0\stackrel{\vec{R}}{\Longrightarrow}\!\!\! ^{\ast}\ C$. 
\end{definition}

We end this section with a simple example, that we will also use in the rest of the paper. 
Consider the following system with a unique membrane:
$$\Pi = (\{a, b, c,\}, [_1\ ]_1, ab, 
\{r_1 = \evrule{a}{b}{}{}, r_2 = \evrule{b}{c}{}{}, r_3 = \evrule{b}{a}{}{})$$
At the initial configuration it is possible to apply one occurrence of the rule $r_1$ and either the rule $r_2$ or the rule $r_3$. If the rule $r_1$ and $r_3$ are applied we obtain the same configuration $ab$, if we apply the rules $r_1$ and $r_2$ we obtain the configuration $bc$. Here only two rules are applicable: $r_2$ or $r_3$ but not both. In the case the latter is used we reach the configuration $cc$, otherwise we use $r_3$ reaching the configuration $ac$ where only $r_1$ can be applied and its application gives the configuration $bc$ again.
In this example each second occurrence of a rule depends on previous ones, for instance if we apply the rules $r_1$ and $r_3$ at the initial configuration, then the second occurrence of $r_1$ and $r_3$ depends on $r_3$ and $r_1$, respectively. The simultaneous occurrence of rules depends on the configuration: at the configuration $ab$ the sets $\{r_1, r_3\}$ and $\{r_1, r_2\}$ are simultaneous, thus $r_1$ can be simultaneous with $r_2$ or $r_3$ but not with both.

\section{Zero-safe Petri nets}\label{sec:net}
A net is a tuple $N = (S, T, F, m_0)$ where $S$ are {\em places}, $T$ are transitions, $F : (S\times T) \cup (T\times S)\rightarrow \mathbb{N}$ is a flow relation and $m_0 : S \rightarrow \mathbb{N}$ is the initial marking.
The evolution of a net is described as usual with the token game.
Let $m : S \rightarrow \mathbb{N}$ be a marking of a net, a finite multiset $U : T \rightarrow \mathbb{N}$ of transitions is {\em enabled} under $m$ if for all $s\in S$ $\sum_{t\in T}U(t)\cdot F(s,t) \leq m(s)$ and the reached marking is $m'(s) = m(s) + \sum_{t\in T}U(t)\cdot (F(t,s) - F(s,t))$, for all $s\in S$.
We then write $m \trans{U} m'$, and call $U$ a \emph{step}. 
A {\em step firing sequence} is defined as follows:
  $m_{0}$ is a step firing sequence, and 
  if $m_{0}\trans{U_{1}}m_{1}\trans{U_{2}}m_{2}\dots 
        m_{n-1}\trans{U_{n}} m_{n}$ is a step firing sequence and
        $m_{n}\trans{U_{n+1}} m_{n+1}$, then 
        $m_{0}\trans{U_{1}}m_{1}\trans{U_{2}}m_{2}\dots 
        m_{n-1}\trans{U_{n}} m_{n} \trans{U_{n+1}} m_{n+1}$ is a step
        firing sequence.  
A marking $m$ is reachable if there is a step firing sequence 
$m_{0}\trans{U_{1}}m_{1}\trans{U_{2}}m_{2}\dots m_{n-1}\trans{U_{n}} m_{n}$ and $m = m_{n}$.
Given a step $U$ and a marking $m$, with $m\trans{U}$ we indicate that $U$ is enabled under $m$ and that there exists a marking $m'$ such that $m\trans{U}m'$. 
A net $(S, T, F, m_{0})$ is \emph{safe} if all the reachable markings are sets and $F(x,y)\leq 1$ for all $x,y\in S\cup T$. 
With $\pre{x}$ ($\post{x}$, respectively) we indicate the multiset $F(\underline{~~},x)$ 
($F(x,\underline{~~})$, respectively).

To be able to represent partial configurations, relevant in understanding causality relations among rule occurrences, we consider zero-safe nets of Bruni and Montanari \cite{BM:TransZSnet}. In these nets the set of places is partitioned into two disjoint sets, the one of \emph{stable} places and the one of zero-safe places.
The intuition is that when a zero-safe place is marked, then the state of the system is \emph{unstable}, meaning that other transitions have still to change the state to reach a \emph{stable} state.
Thus zero safe places can be used to coordinate and synchronize in a single transaction any number of transitions in the net.
The convention we use to draw zero safe places is the usual one: they are represented with smaller circles with respect the ordinary (stable) places. 
\begin{definition}
 A \emph{Zero safe Petri net} (ZS net) is a 
 tuple $N = (S, T, F, m, Z)$ where
 \begin{enumerate}
  \item $N_{\mathit{s}} = (S, T, F, m)$ 
        is a Petri net (the \emph{support}), 
  \item $Z \subset S$ is a subset of places, called \emph{zero safe} places, and 
        $S\setminus Z$ are the \emph{stable} places, and
  \item for all $z\in Z$, $m(z) = 0$.
 \end{enumerate}
 A marking $m'$ is said \emph{stable} iff $m'(z) = 0$ for all $z\in Z$. 
 
 Let $m\trans{U_1}m_1\trans{U_2}m_1 \dots m_{n-1}\trans{U_n}m'$ be a 
 step firing sequence of $N_{\mathit{s}}$, $U = \sum_{i=1}^{n} U_{i}$ is a 
 \emph{stable step} from $m$ to $m'$ if:
 \begin{itemize}
 \item $\forall s \in S\setminus Z$
        $\sum_{t\in T} U(t)\cdot F(s,t) \leq m(s)$, and
 \item $m, m'$ are stable markings.
 
 \end{itemize}
A stable step firing sequence is a step firing sequence where each step is a stable step.
\end{definition}
In a stable step, transitions consuming and producing tokens in zero safe places can fire any number of times provided that their \emph{stable} enabling (i.e., the enabling conditions involving only stable places) is verified before starting the sequence. 
Another notion which will be useful in the following is the one of \emph{stable transaction}, which intuitively capture the idea that all the tokens in the safe places are used in a stable step and the markings produced to reach the final stable one are not stable. For
our purpose we weaken this notion by dropping the requirement that all the tokens in stable places are used but requiring that the tokens left in each stable place are not enough to fire a transition:

\begin{definition}\label{def:stabletrans}
 Let $N = (S, T, F, m, Z)$ be a ZS net and let 
 $m\trans{U_1}m_1\trans{U_2}m_2 \dots m_{n-1}\trans{U_n}m'$ be a 
 stable step, then $U = \sum_{i=1}^{n} U_{i}$ is a 
 \emph{stable transaction} from $m$ to $m'$ if:
 \begin{itemize}
 \item $\forall 1\leq i\leq n-1$, the markings $m_i$ are not stable, and 
 \item $\forall s \in S\setminus Z$, for all $t\in T$, 
       $F(s,t) > m(s) - \sum_{t\in T} U(t)\cdot F(s,t)$.
 \end{itemize}
\end{definition} 
To be able to observe causality, we focus on the so called \emph{non sequential} semantics of a net, where the causal dependencies between transition can be better perceived with respect to the step firing sequence behaviour. We will do so by constructing an unfolding of the Zero-safe net.
We first recall some definitions.

\begin{definition}\label{def:stateofanet}
   Let $N$ be a ZS net, the {\em state} of a net is 
   any finite multiset 
   $X$ of transitions with the property that the function 
   $m_X : S \rightarrow \mathbb{Z}$ given 
   by $m_X(s) = m(s) + \sum_{t\in T}X(t)\cdot (F(t,s) - F(s,t))$, for all $s\in S$, 
   is a reachable marking of the net.
\end{definition}
The notion of state has been introduced in \cite{GP:CS} to characterize 1-unfolding. We use it here to relate computations in membrane systems to reachable markings in the unfolding we will introduce later in this section. 

An occurrence net is a net such that each state is a set and such that a suitable partial order can be associated to it. Formally:

\begin{definition}
\label{one-occnetzs}
  An \emph{occurrence net} $C = \langle B, E, F, m, Z)$ is a safe net satisfying the
  following restrictions:
  \begin{itemize}
  \item  $\forall b\in m$, $\pre{b} = \emptyset$,
  \item  $\forall b\in B$. $\exists b'\in m$ such that $b' F^{*} b$,
  \item  $\forall b\in B$. $|\pre{b}|\leq 1$,
  \item  $F^{+}$ is irreflexive and, for all $e\in E$, the set $\{e'\ |\ e' F^{*} e\}$ is finite,
       and 
  \item  $\#$ is irreflexive, where  
      $e \#_i e'$ iff $e, e' \in E$, $e\neq e'$ and $\pre{e}\cap\pre{e'}\neq \emptyset$, 
      and
      $x \# x'$ iff $\exists\ y, y'\in B\cup E$ such that $y \#_i y'$, $y F^{*} x$ and 
      $y' F^{*} x'$.
  \end{itemize}
 \end{definition}
 On occurrence nets it is easy to define a relation expressing \emph{concurrency}: two elements of
 the causal net are concurrent if they are neither causally dependent nor in conflict. Formally
 $x\ \mathit{co}\ y$ iff $\neg (x \# y$ or $x F^{+} y$ or $y F^{+} x)$. This relation can be 
 extended to sets of conditions: let $A \subseteq B$, then $\mathbf{co}(A)$ iff 
 $\forall b, b'\in A.\ b\ \mathit{co}\ b'$ and 
 $\{e\in E\ |\ \exists b\in A.\ e F^{*} b\}$ is finite.

 We recall now the notion of morphism between nets, which we will use to construct 
 an unfolding. Basically this notion will tell us how to \emph{fold} the occurrence net we are going 
 construct later on the original net.
 \begin{definition}
   \label{de:net morphism}
   Let $N_0$ and $N_1$ be nets.  A {\em morphism} $h: N_0
   \rightarrow N_1$ is a pair $h = (\eta, \beta) $, where
   $\eta : T_0 \rightarrow T_1$ is a partial function and $\beta : S_0 \to
   S_1$ is a multirelation such that 
   (a)    $\mu \beta(m_0) = m_1$ and
   (b)
    for each $t \in T$,
     $\mu \beta(\pre{t}) = \pre{\eta(t)}$, and
     $\mu \beta(\post{t}) = \post{\eta(t)}$.   
 \end{definition}
 
 We can now construct the unfolding of a net.
  \begin{proposition}
 \label{prop:win_unfolding_individual}
  The unfolding $\mathcal{U}(N) = (B, 
  E, F, m, Z_B)$ 
  of the net
  $N = (S, T, F, m, Z)$ is the unique occurrence net to satisfy:
  $$\begin{array}{lcl}
   B  &\!\! =\!\! & \{(m,s,i)\ |\ s\in S\ \mbox{and}\ 0\leq i < m(s)\}\ 
      \bigcup\ \{(\{e\},s,i)\ |\ e\in E\ \mbox{and}\ s\in S\ \mbox{and}\ 
             0\leq i < F_{pre}(\eta(e),s)\} \\ [2mm]
   E  &\!\! =\!\! & \{(X,t)\ |\ X\subseteq B\ \mbox{and}\ \mathbf{co}(X)\ 
            \mbox{and}\ \pre{t} = \mu \beta(X)\} \\ [2mm]
   F &\!\! =\!\! & \left\{\begin{array}{l} F_{pre}((X,t),b)\ \mathit{iff}\ b\in X \\ [1mm]
           F_{post}((X,t),b)\ \mathit{iff}\ \exists s\in S, i\in \mathbb{N}.\ 
            b = ((X,t),s,i) 
            \end{array}\right. \\             [2mm]
   m  &\!\! =\!\! & \{(m,s,i)\ |\ (m,s,i)\in 
            B\}\ 
            \\ [2mm]
   Z_B  &\!\! =\!\! &  \{(m,s,i)\ |\ s\in Z\ \mbox{and}\ 0\leq i < m(s)\}\ 
      \bigcup\ \{(\{e\},s,i)\ |\ e\in E\ \mbox{and}\ s\in Z\ \mbox{and}\ 
             0\leq i < F_{pre}(\eta(e),s)\} \\         
  \end{array}$$
  where $\mathbf{co}$ is the concurrency relation obtained by $F'$ on $B$ and $E$.
  Furthermore $\eta : E \rightarrow T$ defined as 
  $\eta(X,t) = t$ and 
  $\beta : B \rightarrow S$ defined as 
  $\beta(X,s,i) = s$ form a net 
  morphism, called the folding morphism.
 \end{proposition}

Using the following proposition (expressing the fact the reachable markings are
preserved by net morphism) we have that each reachable marking of the unfolding is a reachable marking
of the net.
\begin{proposition}
   \label{pr:net-token-game}
   Let $N_0$ and $N_1$ be nets, and let $h = \langle \eta, \beta
   ): N_0 \rightarrow N_1$ be a net morphism.  For each
   $M, M' \in \mathcal{M}_{N_0}$ and $A \in \mu T$, if
     $M \trans{A} M'$ then $\mu \beta(M) \trans{\mu
       \eta(A)} \mu \beta(M')$.
   Therefore net morphisms preserve reachable markings, i.e. if $M_0$ is a
   reachable marking in $N_0$ then $\mu \beta(M_0)$ is reachable in
   $N_1$.
 \end{proposition} 

The previous proposition tell us that to each reachable marking in the unfolding of a net $N$, a reachable
marking of $N$ corresponds. The following one instead establishes a correspondence among reachable marking of $N$ and states of its unfolding.

\begin{proposition}\label{prop:markingarestate}
   Let $N$ be a ZS net and  $\mathcal{U}(N)$ its unfolding. 
   Let $m_n$ be a reachable marking of $N$. Then there 
   exists a state $X$ of $\mathcal{U}(N)$ 
   such that $m_{n} = \mu\beta(m_X)$. 
\end{proposition}

\section{Event structure}\label{sec:ees}
We briefly recall that a \emph{prime event structure} (\cite{Win:ES}) is the triple
$\mathbf{E} = (E, \leq, \#)$ such that $\leq$ is a partial order and $\#$ is an irreflexive 
and symmetric conflict relation such that $e \# e'$ and $e'\leq e''$ implies 
$e \# e''$ (the so called conflict hereditary principle). We call prime event structure PES.

With $e\ \mathit{co}\ e'$ we 
indicate that $e$ and $e'$ are potentially concurrent, i.e., $\neg e\# e'$ and 
$e \not\leq e'$ and $e'\not\leq e$, and with $\mathbf{co}(X)$ we indicate that all the events in $X$ are potentially pairwise concurrent. 
We recall that a subset $X$ of events is 
\emph{conflict-free} iff $\forall e, e' \in X$ it holds that $\neg e\# e'$.
A configuration of a prime event structure $\mathbf{E} = (E, \leq, \#)$ is any subset $X$ 
of events in $E$ which is conflict-free and $\leq$-closed, i.e.,
$\forall e\in X$, $e'\leq e$ implies $e'\in X$. The set of configurations of a prime event structure is
denoted with $\mathcal{C}_{\mathit{pes}}(\mathbf{E})$.

We introduce now a small variant of the notion of prime event structure. The idea is to add some
information on subset of events that have to occur simultaneously.
\begin{definition}
  Let $\mathcal{A}ct$ be a set of labels. 
  An \emph{event stuctures with simultaneity} (ESS)
  is the tuple $\mathcal{E} = (E, \leq, \#, \mathcal{S}im, \mathcal{L})$ where
  \begin{itemize}
    \item $(E, \leq, \#)$ is a PES, and
    \item $\mathcal{S}im \subseteq \Powfin{E}$ is such that:
      \begin{itemize}
         \item $\emptyset \not\in \mathcal{S}im$ and $\bigcup_{s\in \mathcal{S}im} s = E$, 
         \item $\forall s\in \mathcal{S}im$, $\forall e, e'\in s$, 
               $e\neq e' \Rightarrow e\ \mathbf{co}\ e'$,
         \item $\forall s, s'\in \mathcal{S}im$, $s \cap s'\neq\emptyset$ implies
               that $s\not\subseteq s'$ and $s'\not\subseteq s$,      
         \item $\forall s, s'\in \mathcal{S}im$, $s\cap s'\neq \emptyset$
               implies that $\forall e\in s\setminus s'$, $\forall e'\in s'\setminus s$
               it holds that $e\# e'$, and
      \end{itemize}
   \item $\mathcal{L} : E \rightarrow \mathcal{A}ct$ is a labeling function.          
  \end{itemize}
\end{definition}
The unique novelty in this definition with respect to the usual one of event structure is
that we explicitly indicate which subsets of events may occur simultaneously ($\mathcal{S}im$).
The requirements we pose on $\mathcal{S}im$ are rather obvious: each subset of simultaneous events must
contain only concurrent events, two subsets of simultaneous events may overlap but when they do so then the elements not in common must be in conflict. The last requirement captures the idea that if an event can be simultaneous with two other different events, then these two must belong to alternative computations.
The labeling mapping will play a role when we will discuss the relationship with membrane systems.

\begin{definition}
  Let $\mathcal{E} = (E, \leq, \#, \mathcal{S}im, \mathcal{L})$ be an ESS. Then $X\subseteq E$ is a
  \emph{configuration} iff 
  \begin{itemize}
    \item it is conflict-free,
    \item it is $\leq$-closed, and
    \item there exists a subset $\mathcal{S}'$ of $\mathcal{S}im$ such that 
          $\bigcup_{s\in \mathcal{S}'} s = X$ and 
          $\forall s, s' \in\mathcal{S}'$, $s\cap s' = \emptyset$.
  \end{itemize}
  The set of configurations of an event structure with simultaneity is denoted with 
  $\mathcal{C}_{\mathit{ess}}(\mathcal{E})$.
\end{definition}
Whereas the first two conditions are the usual one for a configuration of an event structure,
the last one simply says that there is a partition of events of a configuration in subsets such that each of this subset is a set of concurrent events.

\begin{proposition}
  Let $\mathcal{E} = (E, \leq, \#, \mathcal{S}im, \mathcal{L})$ be an ESS. Then $\mathbf{E} = (E, \leq, \#)$ 
  is a PES and 
  $\mathcal{C}_{\mathit{ess}}(\mathcal{E})\subseteq \mathcal{C}_{\mathit{pes}}(\mathbf{E})$.
\end{proposition}
The inclusion depends on the fact that some configurations may be ruled out if they do not satisfy the simultaneity requirements.

The vice versa holds as well.

\begin{proposition}
  Let $\mathbf{E} = (E, \leq, \#)$ be a PES. Then 
  $\mathcal{E} = (E, \leq, \#, \mathcal{S}im, id)$ 
  is an ESS, where $\mathcal{S}im = \{\{e\}\ |\ e\in E\}$. Furthermore 
  $\mathcal{C}_{\mathit{ess}}(\mathcal{E}) = \mathcal{C}_{\mathit{pes}}(\mathbf{E})$, and
  $id : E \rightarrow E$ is the identity mapping.
\end{proposition}
In this case it is obvious that the sets of configurations coincide.

\section{From membrane systems to Petri nets}\label{sec:psystandnet}
In this section we recall how to associate a membrane system to a zero safe Petri net and then how evolutions in membrane systems and those in a zero safe nets are related. We follow closely what has been developed by \cite{KK:TCS2008}, adapting it to our setting. To each rule we associate a transition (which are indexed by the \emph{name} of the rule and by the compartment), whereas places are associated to objects. In particular to each object and each membrane we associate two places, one of them being zero safe, connected by a transition consuming tokens in the zero safe place and producing them in the other one (these transitions are denoted with $t^{\hbar}_{(a,i)}$). The zero safe places are used to represent the second component of a partial configuration. The heating of a partial configuration is performed by firing the transitions $t\!$ $^{\hbar}_{q}$, which can be done in a stable step. Finally, the number of tokens in a place gives the number of objects in a membrane.

\begin{definition}\label{de:fromPtoZSL}
Let $\Pi = (V,\mu, w_{1}^{0},\dots,w_{n}^{0},R_{1},\dots,R_{n})$ be a membrane system, then we associate to it the structure 
$\membtonet(\Pi) = (S, T, F, m, Z)$ where:
\begin{itemize}
\item $S = V\times (\{1,\dots,n\}\times\{nz,z\})$, 
      $Z = V\times (\{1,\dots,n\}\times\{z\})$, and
      $T = \bigcup_{i=1}^{n} \{t_i^r \ | \ r\in R_i\}\cup \{t^{\hbar}_{(a,i)} \ | \ a\in V 
      \mbox{ and } 1 \leq i\leq n\}$,
\item for all transitions $t= t_i^r\in T$, with 
      $r = \evrule{u}{v}{}{}$, we define
      
      $F(s,t) = \left\{\begin{array}{ccl}
        u(a) & \quad & \mbox{if}\ j=i\ \mbox{and}\ s = (a,(j,nz))\\
        0 & & \mbox{otherwise} \\
        \end{array}\right.$
        
      $F(t,s) = \left\{\begin{array}{ccl}
        v((a,\mathit{here})) & \quad  & \mbox{if}\ j=i\ \mbox{and}\ s = (a,(j,z))\\
        v((a,\mathit{out})) & & \mbox{if}\ 
                  j = \mathit{father}(i)\ \mbox{and}\ s = (a,(j,z))\\
        v((a,\mathit{in}_j)) & & \mbox{if}\
                  j \in \mathit{children}(i)\ \mbox{and}\ s = (a,(j,z))\\          
        0 & & \mbox{otherwise} \\
        \end{array}\right.$
       
\item for all transitions $t = t^{\hbar}_{(a,i)}\in T$, we define

      $F(s,t) = \left\{\begin{array}{ccl}
       1 & \quad & \mbox{if}\ s = (a,(i,z))\ \mbox{and}\ t = t^{\hbar}_{(a,i)}\\
        0 & & \mbox{otherwise} \\
        \end{array}\right.$
      
      $F(t,s) = \left\{\begin{array}{ccl}
       1 & \quad & \mbox{if}\ s = (a,(i,z))\ \mbox{and}\ t = t^{\hbar}_{(a,i)}\\
        0 & & \mbox{otherwise} \\
      \end{array}\right.$

\item $m(s) = \left\{\begin{array}{ccl}
      w_i(a) & \quad & \mbox{if}\ s = (a,(i,nz)) \\
      0 & & \mbox{otherwise} \\
      \end{array}\right.$
\end{itemize}
\end{definition}

As we said before, the main difference with respect to other approaches is that we add a zero safe place corresponding to each object, playing the r\^{o}le of the second multiset in partial configurations (representing the objects produced while an evolution step is going on), and correspondingly we introduce a number of transition to \emph{heat} to a stable marking. As zero safe places are used to synchronize transitions, we believe that this is the appropriate notion to capture the simultaneity in rule applications of membrane systems.

We illustrate the construction by showing what happens in the case of the rule belonging to
the set or rules associated to the membrane $i$, 
$r = \evrule{aa}{(b,\mathit{here})(c,\mathit{out})(c,\mathit{out})(a,\mathit{in_j})}{}{}\in R_i$.
We assume that $\mathit{father}(i) = k$ and that $\mathit{father}(j) = i$. We draw only places and arcs associated to the transition associated to $r$ which we denote with $t^{r}_i$ (if  the weights is $1$ then the indication of the weight is omitted).
\bigskip

{\footnotesize
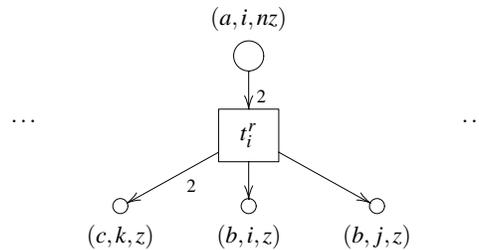
\[
\begin{array}{ccc}
\mbox{\raisebox{-1.6cm}{$\dots$}}\quad
&
\xymatrix@R=5mm@C=12mm{
       &   &    \\
       & \nplace{(a,i,nz)}{}{0cm}{.5cm}\ar[d]^{2}  &
        \\
      & \trs{t^{r}_i}\ar[dr]\ar[d]\ar[dl]^{2} &  \\
    \zplace{(c,k,z)}{}{0cm}{-0.4cm} &  \zplace{(b,i,z)}{}{0cm}{-0.4cm} &
      \zplace{(b,j,z)}{}{0cm}{-0.4cm} \\  
    &  &       \\
}
&
\quad\mbox{\raisebox{-1.6cm}{$\dots$}}
\end{array}
\]
}
\bigskip

\noindent
The places in the first line ($(a,i,nz)$) correspond to the objects in the membrane $i$ (here we assume that the objects are 
$\{a,b,c\}$, and two tokens from the place $(a,i,nz)$ are consumed.
The zero safe places in the bottom line are those that will receive tokens produced by the transition $t^{r}_i$ (the indexes $k, i$ and $j$, in the zero safe places $(c,k,nz), (b,i,nz)$ and $(b,j,nz)$, denote the membrane). The tokens in the zero safe places are removed by the heating transitions.

The following proposition states that the construction in definition \ref{de:fromPtoZSL} gives indeed a ZSL net.
\begin{proposition}
Let $\Pi$ be a membrane system, then $\membtonet(\Pi)$ is a ZS net.
\end{proposition}

\begin{figure}[ht]
{\footnotesize 

\[
\xymatrix@R=10mm@C=6mm{
     &  &  &  &  &  &   &  & \\       
      \nplace{(a,1,nz)}{\bullet}{.75cm}{.4cm}\ar[drr] &  &  &  &
      \nplace{(b,1,nz)}{\bullet}{.5cm}{.5cm}\ar[drr]\ar[drrrr] &  &  &  &  &  &  
      \nplace{(c,1,nz)}{}{.75cm}{.4cm}   \\
      \trs{t_{(a,1)}^{\hbar}}\ar[u] &  & \trs{t^{r_1}_1}\ar[d] &  &
      \trs{t_{(b,1)}^{\hbar}}\ar[u] &  & \trs{t^{r_3}_1}\ar@(d,d)[ddllllllu] &  & \trs{t^{r_2}_1}\ar[drr] &  &  
      \trs{t_{(c,1)}^{\hbar}}\ar[u]  \\
      \zplace{(a,1,z)}{}{-.6cm}{.3cm}\ar[u] &  & 
      \zplace{(b,1,z)}{}{-.6cm}{.3cm}\ar[urr] &  &  &  &  &  &  &  &  
      \zplace{(d,1,z)}{}{.6cm}{.3cm}\ar[u] \\
       &  &  &  &  &  &      \\  
      }
      \]
}
\caption{The ZSI net corresponding to the 
membrane system $\Pi_1$ of section~\ref{psyst}.}\label{fig:netpi}
\end{figure}
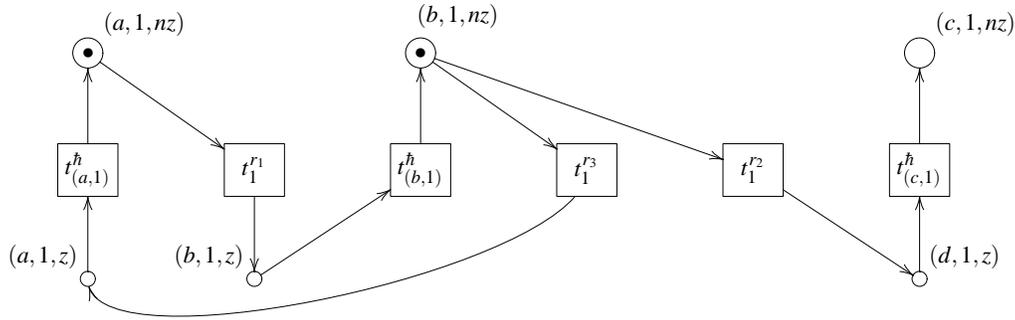

The zero safe net corresponding to the example in section~\ref{psyst} is the one shown in Fig.~\ref{fig:netpi}.

The correspondence between partial configurations and markings is given by the following definition.
\begin{definition}
  Let $\Pi$ be a membrane system, and let
  $\membtonet(\Pi)$ the associated ZS net.
  Let $C = ((w_1,\overline{w}_1), \dots,$ $(w_n,\overline{w}_n))$ be a partial configuration. 
  Then the corresponding 
  marking, denoted with $\nu(C)$, is given, for all $a$ and $i$, 
  by $\nu(C)(a,(i,nz)) = w_i(a)$ and 
  $\nu(C)(a,(i,z)) = \overline{w}_i(a)$.
\end{definition}
With abuse of notation, given a configuration $C = (w_1,\dots, w_n)$, we write $\nu(C)$ for 
$\nu((w_1,\zero),\dots, (w_n,\zero))$.

Evolution steps in membrane system and in the corresponding net are related, as stated in the following propositions. We first show that the effect of a rule and the one of the application of the corresponding transition are related, then we prove that each unstable marking corresponding to a partial configuration can be heated to a stable marking, and finally we prove that to a step in the membrane system a stable step in the ZS net corresponds. 
\begin{proposition}\label{pr:effectscorrespond}
	Let $\Pi$ be a membrane system, and let $\membtonet(\Pi)$ the associated  ZS net.
	Let $C = ((w_1,\overline{w}_1),\dots,$ $(w_n,\overline{w}_n))$ be a partial configuration and 
	$\nu(C)$ the associated marking. If $C \stackrel{\{r,i\}}{\longmapsto} C'$, then 
	$\nu(C)(s)+F(t_i^{r},s)-F(s,t_i^{r}) = \nu(C')(s)$ for all $s\in S$, where $t_i^{r}$ is the transition 
	associated to $r$.
\end{proposition}

\begin{proposition}\label{prop:heatZS}
  Let $\Pi$ be a membrane system and let $\membtonet(\Pi)$ the associated ZS net.
  Let $C = ((w_1,\overline{w}_1),\dots,$ $(w_n,\overline{w}_n))$ be a partial configuration and  
  $\nu(C)$ the associated marking.
  Let $U_{\mathit{heat}}$ be the step where $U_{\mathit{heat}}(t^h_{(a,i)}) = \overline{w}_i(a)$.
  Then $\nu(C)\trans{U_{\mathit{heat}}}m' = \nu(\mathit{heated}(C))$. 
\end{proposition}

Let $\vec{R} = (\widehat{R}_1,\dots,\widehat{R}_n)$ be a vector multi-rule for the membrane system $\Pi$ and let $\membtonet(\Pi)$ the associated ZS net. With $U_{\widehat{R}_i}$ we denote the multiset $U_{\widehat{R}_i}(t_i^{r_i^j}) = \widehat{R}_i(r_i^j)$. With $U_{\vec{R}}$ we denote the
step $\sum_{i=1}^{n}U_{\widehat{R}_i}$.
\begin{proposition}\label{prop:stepcorrespond}
	Let $\Pi$ be a membrane system, and let $\membtonet(\Pi)$ 
	the associated  ZS net.
	Let $C = (w_1,\dots,w_n)$ be a configuration configuration of $\Pi$ and 
	$\nu(C)$ the associated marking of $\membtonet(\Pi)$. 
	If $C \stackrel{\vec{R}}{\Longrightarrow} C'$, then 
         $\nu(C)\trans{U_{\vec{R}}}$, and 
	     there exists a marking $m'$ such that
	      \begin{enumerate}
	         \item 
	            $\nu(C)\trans{U_{\vec{R}}}m'\trans{U_{\mathit{heat}}}\nu(C')$, and
	         \item 
	            $U_{\vec{R}}\oplus U_{\mathit{heat}}$ is a stable transaction.
	      \end{enumerate}
\end{proposition}

Thus to the evolutions in a membrane system stable step firing sequences in the associated net correspond. 
Hence the net mimics the possible computations of the membrane system. 
The corresponding evolutions in the membrane system are obtained by forgetting the transitions emptying the zero safe places.

Consider again the membrane system $\Pi_1$ of section~\ref{psyst}. We have the following evolution (among others): the first micro steps are $(ab,\zero) \stackrel{\{r_1\}}{\longmapsto} (b,b) \stackrel{\{r_3\}}{\longmapsto} (\zero,bc)$ and then $(\zero,bc)$ is heated yielding the configuration $bc$. The second micro step could be $((bc,\zero)) \stackrel{\{r_3\}}{\longmapsto} (c,c)$ and no other rule is applicable. Hence $(c,c)$ is heated and we have $(cc,\zero)$. Thus we have
$(ab) \stackrel{\{r_1,r_3\}}{\Longrightarrow} (bc) \stackrel{\{r_2\}}{\Longrightarrow} (cc)$.
To these micro steps, in the net associated to $\Pi_1$, we have the firing of the corresponding transitions 
($t_1^{r_1}, t_1^{r_3}, t_1^{r_2}$ and the heating ones). It is worth to notice that $t_1^{r_1}, t_1^{r_3}, t_{(b,1)}^{\hbar}, t_{(c,1)}^{\hbar}$ is a stable transaction, as well as $t_1^{r_3}, t_{(c,1)}^{\hbar}$, which are the two steps in the net above.

Using the propositions \ref{pr:effectscorrespond}, \ref{prop:heatZS}  and \ref{prop:stepcorrespond} we have the following theorem.
\begin{theorem}\label{th:conf_are_marking}
	Let $\Pi$ be a membrane system, and let $\membtonet(\Pi)$ be 
	the associated  ZS net.
	Let $C$ be a reachable configuration of $\Pi$. Then 
	$\nu(C)$ is a reachable stable marking of $\membtonet(\Pi)$. 
	Furthermore $\nu(C)$ is reachable with stable transactions.
\end{theorem}

The converse holds as well, when we consider stable step firing sequences where each step is a stable transaction.
\begin{proposition}\label{pr:zerosafeandmembranecorrespond}
	Let $\Pi$ be a membrane system, and let $\membtonet(\Pi)$ be 
	the associated  ZS net. 
	Let $C = (w_1,\dots,$ $w_n)$ be a configuration and 
	$\nu(C)$ the associated marking. 
	Let $\nu(C)\trans{U}m'$ be a stable transaction.
	Then there exists a vector multi-rule
	$\vec{R} =$ $(\widehat{R}_1,\dots,\widehat{R}_n)$ such that 
	$C \stackrel{\vec{R}}{\Longrightarrow} C'$ using $\vec{R}$ and $\nu(C') = m'$.
\end{proposition}

We can state the following theorem.
\begin{theorem}
	Let $\Pi$ be a membrane system, and let $\membtonet(\Pi)$ be 
	the associated  ZS net.
	Let $m$ be a stable marking of $\membtonet(\Pi)$ reachable with a stable step firing sequence
	where each step is a stable transaction.
	Then there exists a reachable configuration $C$ of $\Pi$ such that 
	$\nu(C) = m$.
\end{theorem}

Thus we have seen that to each membrane system it is possible to associate a zero safe net and to the computations of the membrane systems, stable step firing sequences of the net correspond. Furthermore, restricting the stable step firing sequence, we have seen that also the vice versa holds.

We end this section by stating a property of the nets associated to a membrane system.
\begin{proposition}
 Let $\Pi = (V,\mu, w_{1}^{0},\dots,w_{n}^{0},R_{1},\dots,R_{n})$ be a membrane system, and 
 let $\membtonet(\Pi) = (S, T, F,$ $m, Z)$ the associated  ZS net. 
 Then $\pre{Z}\cap\post{Z} = \emptyset$ and $\pre{Z}\cup\post{Z} = T$.
\end{proposition}
This proposition says that the transitions of the net associated to a membrane systems can be partitioned
into two disjoint subsets: the transitions consuming tokens from a zero safe place and those producing 
tokens in the a zero safe place, the former are the heating transitions whereas the latter are those corresponding to a rule of the membrane systems. 

\section{Event structures with simultaneity for membrane systems}\label{sec:evstruandmemb}
In the previous sections we have seen how to associate a zero safe nets to a membrane system
and we have also seen how to unfold a zero safe net.  
In this section we show how to associate an event structure with simultaneity to the unfolding 
of a ZS net associated to a membrane system.

We recall first a well known proposition relating occurrence nets to event structures. 
\begin{proposition}
 \label{prop:pescorrtounf}
 Let $(B, E, F, m, Z)$ be an occurrence net.
 Then $(E, F^{\ast}, \#)$ is a PES, where $\#$ is the conflict relation
 of definition~\ref{one-occnetzs}.
\end{proposition}
Our problem is to find a way to characterize the simultaneous events among those that are concurrent.
For this we need some more notions. A \emph{slice} of an occurrence net $(B, E, F, m, Z)$ is a subset
$A$ of conditions such that $\mathbf{co}(A)$ and there exists a state $X$ such that $m_X = A$.
A \emph{stable} slice is a slice that correspond to a stable marking. 
Consider now a slice which is not stable, i.e., a slice with some conditions which correspond to zero safe
places, which we call \emph{unstable}. 
An unstable slice $A$ is maximal iff $\pre{(A\cap Z)}\cup\post{(A\cap Z)}$ is a transaction.
Maximal unstable slice are those slice that are reached executing all the possible rule occurrences (and
then heating the zero safe conditions) that can be executed.
This notion is not enough, as the following example shows (we draw a part of the unfolding of the net
corresponding to the running example):

{\footnotesize 

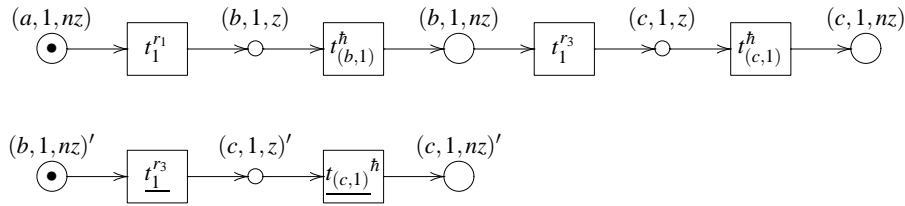
\[
\xymatrix@R=10mm@C=8mm{
    &  &  &  &  &  &   &  & \\       
      \nplace{(a,1,nz)}{\bullet}{0cm}{.4cm}\ar[r] & \trs{t^{r_1}_1}\ar[r]  & 
      \zplace{(b,1,z)}{}{0cm}{.4cm}\ar[r] & \trs{t_{(b,1)}^{\hbar}}\ar[r] &
      \nplace{(b,1,nz)}{}{0cm}{.4cm}\ar[r] & \trs{t^{r_3}_1}\ar[r] & 
      \zplace{(c,1,z)}{}{0cm}{.4cm}\ar[r]  & \trs{t_{(c,1)}^{\hbar}}\ar[r] & 
      \nplace{(c,1,nz)}{}{0cm}{.4cm}  \\
      \nplace{(b,1,nz)'}{\bullet}{0cm}{.4cm}\ar[r] & \trs{\underline{t^{r_3}_1}}\ar[r]  & 
      \zplace{(c,1,z)'}{}{0cm}{.4cm}\ar[r] & \trs{\underline{t_{(c,1)}}^{\hbar}}\ar[r] &
      \nplace{(c,1,nz)'}{}{0cm}{.4cm} &  & 
        &  & 
        \\
      &  &  &  &  &  &  & &     \\  
      }
      \]
}

\noindent consider the slice with the two zero safe places $(c,1,z)$ and $(c,1,z)'$. Certainly 
$\pre{\{(c,1,z),(c,1,z)'\}}\cup\post{\{(c,1,z),(c,1,z)'\}}$ is a transaction, but the point is that this transaction is executed at a marking such that there are two tokens in the place $(b,1,nz)$. 
Hence we add the requirement that the $A' = (A\setminus (A\cap Z))\cup \pre{(\pre{(A\cap Z)})}$ is a marking that is reached with a stable step firing sequence such that each step is a stable transaction.
This new slice is a stable slice, as $\pre{(\pre{(A\cap Z)})}$ are stable places due to the peculiar form of the nets corresponding to membrane system. The slices satisfying these requirements are called 
\emph{maximally simultaneous}.

The following obvious proposition tell us that the events in the preset of a slice are simultaneous.
\begin{proposition}
 Let $\Pi = (V,\mu, w_{1}^{0},\dots,w_{n}^{0},R_{1},\dots,R_{n})$ be a membrane system, and 
 let $\membtonet(\Pi) = (S, T, F,$ $m, Z)$ the associated  ZS net. 
 Let $\mathcal{U}(\membtonet(\Pi)) = (B, E, F, m, Z)$ its unfolding and let $A$ be a maximally simultaneous 
 unstable slice of $\mathcal{U}(\membtonet(\Pi))$. Then $\pre{(A\cap Z)}$ is a set of simultaneous events.
\end{proposition}

Consider now an occurrence net $(B, E, F, m, Z)$ the set $\mathcal{Q} = \{A\ |\ A$ is a maximally simultaneous unstable slice$\}$. Take any $A, A'\in \mathcal{Q}$. 
If $\pre{(A\cap Z)}\cap \pre{(A'\cap Z)}\neq \emptyset$ then it is never be the case that $\pre{(A\cap Z)}$ 
is contained in $\pre{(A'\cap Z)}$ or vice versa, and furthermore the events not in common are in conflict because of the maximality of each step.

We have now all the ingredients we need. We first show that the simultaneous events identified in the previous proposition are those we need to obtain an ESS, and then we state the main results of this section.
\begin{proposition}
 Let $\Pi$ be a membrane system, and 
 let $\membtonet(\Pi) = (S, T, F,$ $m, Z)$ the associated  ZS net. 
 Let $\mathcal{U}(\membtonet(\Pi)) = (B, E, F, m, Z)$ its unfolding and 
 let $\mathcal{Q} = \{A\ |\ A$ is a maximally simultaneous 
 unstable slice$\}$.
 Let $E' = E\cap\pre{Z}$, and let $\mathcal{L} : E' \rightarrow T$ defined as $\eta(e)$ for $e\in E'$,
 where $\eta$ is the mapping associated to the unfolding.
 Then 
 $(E', F^{\ast}\cap (E'\times E'), \#\cap (E'\times E'), \{\pre{(A\cap Z)}\ |\ A\in \mathcal{Q}\}, 
 \mathcal{L})$ 
 is an event structure with simultaneous events.
\end{proposition}
Each event in this event structure is labelled with the transition of which it is occurrence of.
Let $X\in \mathcal{C}_{\mathit{ees}}(\mathcal{E})$ where $\mathcal{E}$ is the EES in the previous proposition, then we associate to it a vector multi-rule by simply counting all the events that corresponds to the same rule. 
\begin{theorem}
 Let $\Pi = (V,\mu, w_{1}^{0},\dots,w_{n}^{0},R_{1},\dots,R_{n})$ be a membrane system, and 
 let $\mathcal{E} = (E, \leq, \#, \mathcal{S}im, \mathcal{L})$ the associated  ESS. 
 Then the followings hold:
 \begin{enumerate}
 \item if $C$ is a reachable configuration and $\vec{R}$ is the associated rule occurrences,  
       then $X$ is a configuration of $\mathcal{E}$ corresponding to $\vec{R}$; and
 \item if $X$ is a configuration of $\mathcal{E}$ there exists a reachable configuration
       $C$ of $\Pi$, that is reached using the rule occurrences  $\vec{R}$ corresponding to $X$.
 \end{enumerate}
\end{theorem}
Indeed we know that to each configuration of the membrane system, a reachable marking in the associated
net correspond and to it a state of its unfolding. By observing that to each state of the unfolding a configuration in the associated event structure corresponds, we have the main result of this paper.

As a consequence of this theorem we have that to each step in a computation in a membrane system a set of simultaneous events in the associated event structure corresponds. For instance, to the computation
$(ab) \stackrel{\{r_1,r_3\}}{\Longrightarrow} (bc) \stackrel{\{r_3\}}{\Longrightarrow} (cc)$ the
configuration $\{t^{r_1}_1, t^{r_3}_1, \underline{t^{r_3}_1}\}$ corresponds, where  
$\{t^{r_1}_1, t^{r_3}_1\}$ and $\{\underline{t^{r_3}_1}\}$ belong to $\mathcal{S}im$ and furthermore
$\mathcal{L}(\{t^{r_1}_1, t^{r_3}_1\}$ gives the transitions corresponding to $\{r_1,r_3\}$.
Thus from each configuration it is possible to determine the corresponding computation in the membrane system.

\section{Conclusions}
In this paper we have presented a new notion of event structures that is able to capture not only the causal dependencies among rule occurrences of a membrane system, but also the simultaneity. 
This notion is a conservative extension of the well know notion of prime event structure of Winskel.

The event structure associated to a membrane system is obtained as follows. First one associates to each membrane system a zero safe net. Following the notion of partial configuration introduced by Nadia Busi, we argued that an evolution step in a P system can be represented as a suitable transaction in a ZSI net, where the zero places are used to \emph{synchronize} the various rules applied in a step. Then the zero safe net is unfolded and some markings of this unfolding are characterized as markings corresponding to simultaneous executions. 
This characterization allows to avoid the so called \emph{barb}-events that are used in 
\cite{KKR:PetriNetsandMembrane2005,KKR:ProcessforMembrane2006} to capture the maximality in step executions of membrane systems.

Our characterization is based on the \emph{reachability} of markings using transactions, but it would be useful to find a structural characterization instead.

As we already pointed in the introduction, the approach we have pursued here is based on the called individual token philosophy, whereas in \cite{PiSa:ASC} we argued that the so called collective token philosophy seizes the causality arising in membrane systems. We plan to extend the notion of event structure with simultaneity to cope with this view, along the lines developed in \cite{GP:CSESPN}.


\end{document}